\documentclass[aps,pra,10pt]{revtex4-2}
\usepackage{amssymb}
\usepackage[utf8]{inputenc}
\usepackage[english]{babel}
\usepackage{amsmath}
\usepackage{amsthm}
\usepackage{mathrsfs}

\newtheorem{proposition}{Proposition}[section]
\newtheorem{theorem}[proposition]{Theorem}
\theoremstyle{remark}
\newtheorem{remark}[proposition]{Remark}

\begin{document}
\raggedbottom
\title{Generalized Kosmann derivative and symmetry operators on Dirac structures}

\author{\"{U}mit Ertem}
\email{umitertem@ankara.edu.tr}
\affiliation{Department of Software Engineering, Ankara University,\\
Faculty of Engineering, 06830 G\"olba\c{s}{\i}, Ankara, T\"urkiye}

\author{\"{O}zg\"{u}r A\c{c}{\i}k}
\email{ozacik@science.ankara.edu.tr}
\affiliation{Department of Physics, Ankara University,\\
 Faculty of Sciences, 06100, Tando\u{g}an-Ankara, T\"urkiye}

\author{\"{O}zg\"{u}r Kelek\c{c}i}
\email{okelekci@thk.edu.tr}
\affiliation{Department of Basic Sciences, Faculty of Engineering,\\
 University of Turkish Aeronautical Association, Ankara, T\"urkiye}

\begin{abstract}
We develop a spinorial symmetry framework for metric Dirac structures
in $H$-twisted exact Courant algebroids. A generalized metric equips a
Dirac structure with an intrinsic Riemannian Lie-algebroid geometry and
identifies it isometrically with the positive subbundle of the
generalized tangent bundle. For sections preserving the generalized
metric, we prove an exact intertwining between the positive-spinor
Kosmann--Dorfman action and the intrinsic Kosmann derivative. The
comparison is independent of the auxiliary compatible generalized
connection. Within the intrinsic geometry we derive Killing--Yano (KY)
and conformal Killing--Yano (CKY) integrability identities, construct
Killing-spinor-preserving operators from odd special KY and even
special closed CKY forms, and obtain these forms from Hermitian
bilinears of Killing spinors with imaginary Killing number. We
illustrate the framework on a global Dirac structure over
$S^2\times S^1$ with nontrivial $H$-flux class and rank-two anchor.
Its spin connection and symmetry operators are explicit, and its
Killing-spinor space is a transverse family
$C^\infty(S^1,\mathbb C^2)$. Finally, intrinsic Killing sections act on Killing spinors by a Lie
algebra representation. For a graph Dirac structure over $S^3$, the odd
Jacobi identity of the Hermitian-current bracket fails by a phase, and
adjoining a single even phase generator yields a Lie superalgebra. These
results connect generalized isometries, hidden spinorial symmetries,
and flux-sensitive Lie-algebroid geometry.
\end{abstract}

\maketitle
\enlargethispage{-\baselineskip}

\section{Introduction}
Generalized geometry brings diffeomorphisms and differential-form
transformations into a common geometric language on
$E=TM\oplus T^*M$, and its Courant-algebroid description makes the
role of closed three-form flux particularly transparent
\cite{Hitchin,Gualtieri,Roytenberg,CoimbraConstableWaldram}.
A Dirac structure is an involutive maximally isotropic subbundle of
$E$, hence a Lie algebroid in its own right \cite{Courant}.
Its anchor need not be invertible: a Dirac structure can therefore
encode symmetries and differential constraints that are not captured
by ordinary Riemannian geometry on the base manifold. Equipping it
with the metric induced by a generalized metric provides a natural
setting in which to investigate spinors, Killing--Yano forms, and
symmetry operators in the presence of flux.

An important feature of this setting is the coexistence of two
Clifford geometries. The Courant pairing restricts to zero on a
Dirac structure $L$, while the induced metric $h_L$ is positive
definite. The intrinsic spin representation of $(L,h_L)$ is therefore
distinct from the ambient generalized-spinor representation. The
generalized metric nevertheless identifies $L$ isometrically with its
positive eigenbundle $V_+$, providing a geometric route for comparing
their spinorial symmetry actions. This comparison becomes especially
interesting for generalized Killing sections, since their Dorfman
action preserves both the Dirac structure and the generalized metric.

The first main result is an \emph{exact positive-spinor comparison
theorem}. For an $H$-twisted Dirac structure, a chosen spin structure,
and a generalized Killing section $\mathcal K\in\Gamma(L)$, we show
that the positive-spinor Kosmann--Dorfman action, transported along the
metric-induced isometry $L\simeq V_+$, equals the intrinsic Kosmann
derivative. The proof does not require choosing an ambient connection
whose restriction is the intrinsic Levi--Civita connection: the
difference between the connections cancels against the algebraic spin
generator. This yields a precise, connection-independent comparison
between two familiar spinorial constructions in a setting where their
relation is not automatic.

The second result develops the corresponding \emph{hidden-symmetry
geometry}. We derive intrinsic Killing--Yano (KY) and conformal
Killing--Yano (CKY) integrability relations directly from Lie-algebroid
curvature, with the correct rank-dependent coefficients. Under the
special-form conditions, odd KY and even closed CKY forms generate
operators preserving intrinsic Killing spinors. A direct Hermitian
bilinear construction supplies these special forms from Killing
spinors with imaginary Killing number and determines their common
constant $c=-4\lambda^2$. The resulting formalism places the
ordinary higher-form symmetry mechanism in a metric Dirac-algebroid
setting, where the bracket and connection can retain nontrivial
flux and anchor geometry.

A central application demonstrates that these structures are not
restricted to graph Dirac structures equivalent to the tangent
bundle. On $M=S^2\times S^1$, we construct a global $H$-twisted
Dirac structure with
$H=-\tfrac12\Omega\wedge dt$, nonvanishing flux period,
and rank-two anchor. Its intrinsic connection has an explicit global
orthonormal frame and admits Killing spinors with
$\lambda=-i/4$. The full solution space is
$C^\infty(S^1,\mathbb C^2)$, exhibiting transverse degrees of
freedom that arise from the noninvertible anchor. Nonzero KY/CKY
symmetry actions and a family of generalized Killing sections are
computed explicitly. Although the flux background is related to a
standard Hopf T-duality construction
\cite{BouwknegtEvslinMathai,CavalcantiGualtieri}, the intrinsic spinorial structures and
their symmetry actions are the objects calculated here.

The geometry also supports a concrete analysis of symmetry brackets.
Intrinsic Killing sections act on Killing spinors by a Lie algebra
representation, and their Hermitian currents transform equivariantly.
On the $S^3$ graph Dirac structure, the simplest current bracket fails
the odd Jacobi identity by a phase rotation. We show that adjoining one
even generator acting by this phase and adjusting the odd bracket
produces a real Lie superalgebra; the same mechanism, with a family
of phase generators, applies on the $S^2\times S^1$ flux background.
This explicit construction clarifies how algebraic closure can be
recovered and motivates the search for analogous extensions in more
general Dirac-algebroid geometries.

Our work builds on the spinorial Lie derivative of Kosmann
\cite{Kosmann}, the generalized Kosmann--Dorfman construction
\cite{CoimbraConstable}, spin and Dirac operators on Lie algebroids
\cite{Boucetta,BalcerzakPierzchalski,TarvijiRezaii}, and established
KY/CKY symmetry methods in spin geometry
\cite{Ertem,AcikErtem,BennCharlton,CarigliaKrtousKubiznak,Ertem2,Semmelmann}.
It is also complementary to generalized-spinor bilinears and
torsionful hidden-symmetry algebras studied in
\cite{AcikErtemKelekci,ErtemKelekciAcik}. The contribution of the
present work is the precise positive-spinor intertwining, the
intrinsic Dirac-algebroid realization of higher-form symmetries,
their verification on a nontransitive, cohomologically
nontrivial flux background, and an explicit phase-extended
Lie superalgebra in the three-sphere example.

Section~\ref{sec:geometry} fixes the Courant and generalized-metric
conventions and states the conventions used throughout.
Section~\ref{sec:ambient} develops the spin lift and its relation to the
Kosmann--Dorfman derivative. Section~\ref{sec:intrinsic} treats intrinsic
curvature and KY/CKY integrability. Section~\ref{sec:spinor-symmetry}
establishes the spinorial comparison, bilinear identities, and symmetry
operators. Section~\ref{sec:example} constructs the two flux examples,
while Section~\ref{sec:algebra} examines symmetry brackets and their
Jacobi properties. Section~\ref{sec:conclusion} summarizes the results
and future directions.

\section{Generalized Geometry}\label{sec:geometry}

For an $n$-dimensional smooth manifold $M$ and its tangent bundle $TM$ and cotangent bundle $T^*M$, one can define a generalized tangent bundle $E$ as $E=TM\oplus T^*M$. Sections of $E$ are called generalized vectors and a generalized vector $\mathcal{X}=X+\xi$ can be written as a sum of a vector field $X\in\Gamma(TM)$ and a 1-form $\xi\in\Gamma(T^*M)$. A bilinear form $\langle\,,\,\rangle$ on $E$ can be defined for generalized vectors $\mathcal{X}=X+\xi$ and $\mathcal{Y}=Y+\eta$ as
\begin{equation}\label{bilinear}
\langle\mathcal{X}, \mathcal{Y}\rangle=\frac{1}{2}\big(i_X\eta+i_Y\xi\big)
\end{equation}
where $i_X$ denotes the interior derivative or contraction with respect to the vector field $X$. For a 2-form $B\in\Omega^2(M)$ and a Riemannian metric $g:TM\rightarrow T^*M$, one can define a generalized metric on $E$ as follows
\begin{equation}\label{gmetric}
\mathcal{G}=\left(\begin{array}{cc}-g^{-1}B & g^{-1}\\ g-Bg^{-1}B & Bg^{-1}\end{array}\right).
\end{equation}
Since $\mathcal{G}^2=I$, $\pm 1$ eigenspaces of $\mathcal{G}$ which are denoted by $V_{\pm}$ give a metric splitting of $E$
\[
E=TM\oplus T^*M=V_+\oplus V_-
\]
corresponding to maximally positive/negative definite subbundles. The generalized metric can be written in terms of the bilinear form defined in \eqref{bilinear} restricted to $V_\pm$ as
\begin{equation}\label{gmetric2}
h_E=\langle\,,\,\rangle|_{V_+}-\langle\,,\,\rangle|_{V_-}.
\end{equation}
Here $\mathcal G$ denotes the orthogonal involution in \eqref{gmetric};
its associated positive-definite bilinear form is
$h_E(\mathcal X,\mathcal Y):=\langle\mathcal G\mathcal X,\mathcal Y\rangle$.
In particular, a
maximally isotropic subbundle for $\langle\,,\,\rangle$ need not be
isotropic for $h_E$.

Lie bracket $[\,,\,]$ of vector fields can be generalized to a bracket operation on $E$ called Courant bracket as $[\,,\,]_C:\Gamma E\times\Gamma E\rightarrow \Gamma E$. For two generalized vectors $\mathcal{X}=X+\xi$ and $\mathcal{Y}=Y+\eta$, Courant bracket is defined as
\begin{equation}\label{eq7}
[\mathcal{X}, \mathcal{Y}]_C=[X,Y]+\mathcal{L}_X\eta-\mathcal{L}_Y\xi-\frac{1}{2}d\left(i_X\eta-i_Y\xi\right)
\end{equation}
where $\mathcal{L}_X$ is the Lie derivative with respect to the vector field $X$ and $d$ is the exterior derivative. Courant bracket is an antisymmetric bracket but it does not satisfy the Jacobi identity and hence does not correspond to a Lie bracket. However, $E$ with Courant bracket admits a Courant algebroid structure \cite{Roytenberg}. In the presence of a closed three-form $H$, whose de Rham class classifies exact Courant algebroids over $M$, the twisted Courant bracket is
\begin{equation}\label{eq10}
[\mathcal{X},\mathcal{Y}]_H=[\mathcal{X},\mathcal{Y}]_C-i_Xi_YH.
\end{equation}
On the other hand, one can also define a bracket operation called Dorfman bracket on $E$ that satisfies the left Leibniz identity but is not antisymmetric:
\begin{equation}\label{eq11}
[\mathcal{X}, \mathcal{Y}]_D=[X,Y]+\mathcal{L}_X\eta-i_Yd\xi.
\end{equation}
With the convention for $H$ in \eqref{eq10}, the corresponding
$H$-twisted Dorfman bracket is
\[
[\mathcal X,\mathcal Y]_{D,H}
=[\mathcal X,\mathcal Y]_D-i_Xi_YH.
\]
When $dH=0$ this defines the $H$-twisted exact Courant algebroid.
In fact, Courant bracket corresponds to the antisymmetrization of the Dorfman bracket
\begin{equation}\label{eq12}
[\mathcal{X}, \mathcal{Y}]_C=[\mathcal{X}, \mathcal{Y}]_D-d\langle\mathcal{X}, \mathcal{Y}\rangle.
\end{equation}
Dorfman bracket is also a generalization of the Lie derivative on differential forms and it is considered as the generalized Lie derivative
\[
\mathbb{L}_{\mathcal{X}}\mathcal{Y}=[\mathcal{X}, \mathcal{Y}]_D
\]
which satisfies the following identities \cite{Baraglia}
\[
\mathbb{L}_{\mathcal{X}}({f\mathcal{Y}})=(\pi(\mathcal{X})f)\mathcal{Y}+f\mathbb{L}_{\mathcal{X}}\mathcal{Y}
\]
\begin{equation}\label{eq13}
[\mathbb{L}_{\mathcal{X}}, \mathbb{L}_{\mathcal{Y}}]=\mathbb{L}_{[\mathcal{X},\mathcal{Y}]_D}
\end{equation}
\[
\mathbb{L}_{\mathcal{X}}(\mathcal{Y}\otimes \mathcal{Z})=\mathbb{L}_{\mathcal{X}}\mathcal{Y}\otimes\mathcal{Z}+\mathcal{Y}\otimes\mathbb{L}_{\mathcal{X}}\mathcal{Z}.
\]
where $f$ is a function and $\mathcal{X}, \mathcal{Y}, \mathcal{Z} \in \Gamma E$.

A generalized connection $\mathbb{D}$ can also be defined on $E$ which is compatible with the bilinear form $\langle\,,\,\rangle$ as
\[
\mathbb{D}:\Gamma E\rightarrow\Gamma(E^*\otimes E)
\]
which satisfies the following identities
\begin{align}\label{eq15}
\mathbb{D}_{\mathcal{X}}(f\mathcal{Y})&=\pi(\mathcal{X})(f)\mathcal{Y}+f\mathbb{D}_{\mathcal{X}}\mathcal{Y}\\
\pi(\mathcal{X})\langle\mathcal{Y},\mathcal{Z}\rangle&=\langle\mathbb{D}_{\mathcal{X}}\mathcal{Y},\mathcal{Z}\rangle+\langle\mathcal{Y},\mathbb{D}_{\mathcal{X}}\mathcal{Z}\rangle
\end{align}
where $f$ is a function, $\mathcal{X},\mathcal{Y},\mathcal{Z}\in\Gamma E$.
We also require $\mathbb D_{f\mathcal X}\mathcal Y
=f\mathbb D_{\mathcal X}\mathcal Y$.
Compatibility with the natural pairing, displayed above, is distinct
from generalized-metric compatibility.  The latter is the additional
condition $\mathbb D\mathcal G=0$, equivalently
\[
\pi(\mathcal X)h_E(\mathcal Y,\mathcal Z)
=h_E(\mathbb D_{\mathcal X}\mathcal Y,\mathcal Z)
+h_E(\mathcal Y,\mathbb D_{\mathcal X}\mathcal Z).
\]
The calculation in Section~\ref{sec:ambient} assumes both
compatibilities and, where explicitly stated, vanishing generalized
torsion. Such compatible torsion-free generalized connections need not be
unique \cite{GarciaFernandezTorsionfree}.
Moreover, generalized torsion 3-form $T_{\mathbb{D}}\in\Gamma (\Lambda^3E^*)$ of a generalized connection $\mathbb{D}$ on $E$ is defined by
\begin{equation}
T_{\mathbb{D}}(\mathcal{X},\mathcal{Y},\mathcal{Z})=\langle\mathbb{D}_{\mathcal{X}}\mathcal{Y}-\mathbb{D}_{\mathcal{Y}}\mathcal{X}-[\mathcal{X},\mathcal{Y}]_D,\mathcal{Z}\rangle+\langle\mathbb{D}_{\mathcal{Z}}\mathcal{X},\mathcal{Y}\rangle.
\end{equation}
This tensor refers to the untwisted Dorfman bracket.  In an
$H$-twisted calculation the corresponding tensor
$T^{(H)}_{\mathbb D}$ is defined by replacing
$[\,,\,]_D$ with $[\,,\,]_{D,H}$ in this formula.
The three-form $H$ enters through the bracket; it is not, in general,
the generalized torsion of $\mathbb D$.

\medskip\noindent\emph{Conventions and scope.}
Throughout, the Courant and Dorfman brackets and all torsion and
integrability statements refer to the same closed three-form $H$.
Compatibility of a generalized connection with the natural pairing,
compatibility with the generalized metric and vanishing generalized
torsion are independent conditions and are imposed separately where
they are used. Three spinor modules occur: the ambient
$\mathrm{Cl}(E)$-module $S_E$ of Section~\ref{sec:ambient}, the
intrinsic spinor bundle $S_L$ of a metric Dirac structure $(L,h_L)$,
and the spin module $S(V_+)$ obtained by transporting the spin
structure of $L$ along the isometry $L\simeq V_+$ of
Section~\ref{sec:intrinsic}. The comparison theorem of
Section~\ref{sec:spinor-symmetry} relates $S_L$ and $S(V_+)$; neither
is identified with $S_E$. Intrinsic differential operators,
curvatures and Killing-type equations are formed with the
Lie-algebroid differential $d_L$, the coderivative $\delta_L$ and the
torsion-free Levi--Civita $L$-connection $\nabla^L$ of the rank-$n$
bundle $L$; hence their dimension-dependent coefficients involve
$n$, not $2n$. The flux enters these intrinsic structures through
the twisted bracket, not as torsion of $\nabla^L$.

\section{Generalized Kosmann Derivative}\label{sec:ambient}

The operator underlying the proposed spinorial action is the difference
between the (twisted) Dorfman derivative and a generalized connection.
We use the same exact Courant algebroid and the same closed three-form
$H$ on both sides of this comparison.  Set
\begin{equation}\label{eq:generalized-A}
\mathcal A_{\mathcal X}\mathcal Y
 :=\mathbb L^{H}_{\mathcal X}\mathcal Y
    -\mathbb D_{\mathcal X}\mathcal Y,
\qquad \mathbb L^H_{\mathcal X}\mathcal Y
  :=[\mathcal X,\mathcal Y]_{D,H}.
\end{equation}
Since $\mathcal A_{\mathcal X}f=0$, the operator
$\mathcal A_{\mathcal X}$ is $C^\infty(M)$-linear in the
argument on which it acts.  Both the Dorfman derivative and
$\mathbb D$ preserve the natural split-signature pairing
$\langle\,,\,\rangle$.  Therefore
\begin{equation}\label{eq:A-skew}
\langle \mathcal A_{\mathcal X}\mathcal Y,\mathcal Z\rangle
+\langle \mathcal Y,\mathcal A_{\mathcal X}\mathcal Z\rangle=0.
\end{equation}
For any $\mathcal X$, the fibrewise endomorphism
$\mathcal A_{\mathcal X}$ thus belongs to
$\mathfrak{so}(E,\langle\,,\,\rangle)$.
The generalized Killing condition is \emph{not} needed for this
statement: if $\mathbb D\mathcal G=0$, then
\begin{equation}\label{eq:A-metric}
(\mathbb L^H_{\mathcal X}\mathcal G)=0
\quad\Longleftrightarrow\quad
[\mathcal A_{\mathcal X},\mathcal G]=0,
\end{equation}
so a generalized Killing vector additionally ensures that
$\mathcal A_{\mathcal X}$ is block-diagonal with respect to
$E=V_+\oplus V_-$.

Let $\{\mathcal F_A\}_{A=1}^{2n}$ be a local frame of $E$ and
$\{\mathcal F^A\}$ its dual frame with respect to the natural
pairing, so that
$\langle\mathcal F^A,\mathcal F_B\rangle=\delta^A{}_B$.
Define the antisymmetric components
\begin{equation}\label{eq:A-components}
\mathcal A_{\mathcal X,AB}
:=\langle\mathcal A_{\mathcal X}\mathcal F_A,\mathcal F_B\rangle
=-\mathcal A_{\mathcal X,BA}.
\end{equation}
The $H$-twisted generalized torsion of $\mathbb D$, as defined
in Section~\ref{sec:geometry}, gives the identity
\begin{equation}\label{eq:A-torsion}
\mathcal A_{\mathcal X,AB}
=-\langle\mathbb D_{\mathcal F_A}\mathcal X,\mathcal F_B\rangle
+\langle\mathbb D_{\mathcal F_B}\mathcal X,\mathcal F_A\rangle
-T^{(H)}_{\mathbb D}(\mathcal X,\mathcal F_A,\mathcal F_B).
\end{equation}
This follows directly by inserting
$\mathbb L^H_{\mathcal X}\mathcal F_A
=\mathbb D_{\mathcal X}\mathcal F_A
+\mathcal A_{\mathcal X}\mathcal F_A$ into the torsion
formula.  In particular, for
$T^{(H)}_{\mathbb D}=0$, the infinitesimal generator is the
antisymmetrized covariant derivative of $\mathcal X$.

\subsection{Spin lift and normalization}

In a splitting, Clifford multiplication on polyforms is
\begin{equation}\label{eq:ambient-clifford}
c_E(X+\xi)\alpha=i_X\alpha+\xi\wedge\alpha,
\qquad \{c_E(u),c_E(v)\}=2\langle u,v\rangle.
\end{equation}
To fix the scalar convention, choose the standard spin structure with
\begin{equation}\label{eq:ambient-density-module}
S_E=\Lambda^\bullet T^*M\otimes |\det TM|^{1/2}\otimes\mathbb C,
\end{equation}
as in \cite[Secs.~2.3, 2.8]{Gualtieri}. This is the spin module,
whereas unweighted polyforms carry the corresponding weighted action.
A positive density $\mu\in\Gamma(|\det T^*M|)$ trivializes the line
by $\mu^{-1/2}$. All spin lifts below are induced from the stated
spin principal bundle, without an additional line-bundle connection.

Assume that $\mathbb D$ is lifted to a connection
$\mathbb D^{S_E}$ on the ambient Clifford module $S_E$,
with $[\mathbb D^{S_E}_{\mathcal X},c_E(\mathcal Y)]
=c_E(\mathbb D_{\mathcal X}\mathcal Y)$.
The standard infinitesimal spin lift of an element of
$\mathfrak{so}(E,\langle\,,\,\rangle)$ is then
\begin{equation}\label{eq:spin-lift}
\mathbb L^{\mathrm{spin}}_{\mathcal X}\psi
=\mathbb D^{S_E}_{\mathcal X}\psi
-\frac14\,\mathcal A_{\mathcal X,AB}
\,c_E(\mathcal F^A)c_E(\mathcal F^B)\psi.
\end{equation}
The factor $1/4$ in \eqref{eq:spin-lift} accompanies a sum over
ordered indices $A,B$ of products of two Clifford vectors. With this
normalization the Clifford anticommutation rule yields
\begin{equation}\label{eq:spin-commutator}
\left[-\frac14\mathcal A_{\mathcal X,AB}
  c_E(\mathcal F^A)c_E(\mathcal F^B),c_E(\mathcal Y)\right]
=c_E(\mathcal A_{\mathcal X}\mathcal Y).
\end{equation}
Consequently,
\begin{equation}\label{eq:spin-intertwines-clifford}
[\mathbb L^{\mathrm{spin}}_{\mathcal X},c_E(\mathcal Y)]
=c_E(\mathbb L^H_{\mathcal X}\mathcal Y).
\end{equation}
For vanishing generalized torsion, the algebraic correction in
\eqref{eq:spin-lift} may equivalently be written as
\begin{equation}\label{eq:torsionfree-spin}
\frac14\bigl(
\langle\mathbb D_{\mathcal F_A}\mathcal X,\mathcal F_B\rangle
-\langle\mathbb D_{\mathcal F_B}\mathcal X,\mathcal F_A\rangle
\bigr)c_E(\mathcal F^A)c_E(\mathcal F^B).
\end{equation}
The Clifford commutator alone determines the lift only up to a
scalar on the irreducible complex ambient Clifford module.
The spin-density convention \eqref{eq:ambient-density-module}
and the induced spin connection fix this scalar.

A particularly useful connection-independent realization exists
on unweighted polyforms in the splitting $E=TM\oplus T^*M$.
With the sign of the twist fixed in Section~\ref{sec:geometry}, write
$d_H=d-H\wedge$; for $dH=0$, $d_H^2=0$.  The graded
anticommutator with Clifford multiplication gives
\begin{equation}\label{eq:polyform-lift}
\mathcal L^{H,\mathrm{poly}}_{X+\xi}\psi
:=[d_H,c_E(X+\xi)]_+\psi
=\mathcal L_X\psi+(d\xi-i_XH)\wedge\psi.
\end{equation}
Directly using $[\beta\wedge,i_Y]=-(i_Y\beta)\wedge$
for a two-form $\beta$ shows that
\begin{equation}\label{eq:polyform-lift-bracket}
[\mathcal L^{H,\mathrm{poly}}_{\mathcal X},c_E(\mathcal Y)]
=c_E([\mathcal X,\mathcal Y]_{D,H}).
\end{equation}
The spin lift on \eqref{eq:ambient-density-module}, expressed in
the trivialization $\alpha\otimes\mu^{-1/2}\leftrightarrow\alpha$, is
\begin{equation}\label{eq:density-correction}
\mathbb L^{\mathrm{spin}}_{X+\xi}
=\mathcal L^{H,\mathrm{poly}}_{X+\xi}
-\frac12\operatorname{div}_{\mu}X,
\qquad \mathcal L_X\mu=(\operatorname{div}_{\mu}X)\mu.
\end{equation}
Indeed, the infinitesimal $\mathfrak{gl}(n)$ action on the spin
module differs from its tensor action on forms by half the trace;
in coordinates the vector-field part has spin generator
$(\partial_jX^i)dx^j\wedge i_{\partial_i}
-\tfrac12\partial_iX^i$. Changing from the coordinate density
to $\mu$ replaces this trace by $\operatorname{div}_{\mu}X$.
The two-form term is already a trace-free Clifford generator.
This also follows by differentiating the density factor
$\mu^{-1/2}$. Changing the auxiliary pairing-compatible connection
adds a spin endomorphism to $\mathbb D^{S_E}_{\mathcal X}$
and its negative to the lift of $\mathcal A_{\mathcal X}$;
the operator is unchanged. For a generalized Killing section,
$X$ is Killing for $g$, so the scalar term vanishes for
$\mu=\operatorname{vol}_g$.

\subsection{Relation to the Kosmann--Dorfman derivative}

The Kosmann--Dorfman (KD) derivative is already defined in the
literature, in particular by Coimbra and Strickland-Constable
\cite{CoimbraConstable}.  Their definition is expressed through
a torsion-free compatible generalized connection and the adjoint
projection for the relevant compact structure group:
\begin{equation}\label{eq:KD-reference}
\mathbb L_V^{\mathrm{KD}}\epsilon
=\mathbb D_V\epsilon
-(\mathbb D\times_{\mathrm{ad}P}V)\cdot\epsilon.
\end{equation}
For the $O(n,n)$ case the relevant generalized-metric reduction
is $\operatorname{Spin}(n)\times\operatorname{Spin}(n)$;
the cited work also treats exceptional-generalized spinors.
Our construction \eqref{eq:spin-lift} is an ambient
$\operatorname{Spin}(n,n)$ lift of the same kind of infinitesimal
Dorfman action; for a generalized Killing vector,
\eqref{eq:A-metric} selects the compact metric-preserving
subalgebra. Equality with a KD operator on another spinor
representation requires the corresponding representation map,
projection and normalization.

For a generalized Killing section the compact action is the spin
lift of the restriction of $\mathcal A_{\mathcal X}$ to $V_+$
or $V_-$. With a torsion-free compatible connection this is the
corresponding adjoint-projection KD action: the torsion identity
\eqref{eq:A-torsion} identifies $\mathcal A_{\mathcal X}$ with
the negative covariant adjoint generator in
\eqref{eq:KD-reference}. Section~\ref{sec:bridge} proves the
comparison with intrinsic spinors for sections in $L$, on the
specified positive spin representation. More explicitly, let $\{v_a\}$ be orthonormal
in $V_+$ for $\langle\,,\,\rangle$. For a generalized Killing
section $V$ and a torsion-free compatible $\mathbb D$,
\begin{equation}\label{eq:KD-positive-components}
\mathbb L^{\mathrm{spin},+}_V
=\mathbb D^{S(V_+)}_V
+\frac14\langle\mathbb D_{v_a}V,v_b\rangle
 [c_+(v_a),c_+(v_b)].
\end{equation}
Indeed, \eqref{eq:A-torsion} gives the antisymmetrized derivative,
and its Clifford contraction gives the commutator displayed above.
Equivalently the correction is one half of the covariant derivative
contracted with the antisymmetrized Clifford product. The symmetric
part of that derivative need not vanish; it is removed by the
Clifford commutator.
This is precisely the positive compact component of
\cite[Eq.~(3.14)]{CoimbraConstable}, with unit positive frames and
antisymmetrized Clifford products. The reduction of KD to the ordinary
Kosmann derivative for generalized Killing vectors is already
established there; Theorem~\ref{prop:bridge} makes the separate
comparison with the intrinsic metric Dirac-algebroid action.

\section{Intrinsic Geometry of a Dirac Structure}\label{sec:intrinsic}

Let $L\subset E$ be a smooth maximally isotropic subbundle for the
natural pairing, so that $\operatorname{rank}L=n$ and
$\langle u,v\rangle=0$ for all $u,v\in\Gamma(L)$.
It is an ($H$-twisted) Dirac structure if
$[\Gamma(L),\Gamma(L)]_{D,H}\subset\Gamma(L)$.
The bracket $[u,v]_{L,H}:=[u,v]_{D,H}$ is then antisymmetric
on $\Gamma(L)$ and, together with the anchor
$\rho_L=\pi|_L$, equips $L$ with a Lie algebroid structure
\cite{Courant,Gualtieri}.  The nondegenerate pairing of $E$ cannot
be used to raise or lower indices intrinsic to $L$, because its
restriction to $L$ vanishes.

The generalized metric instead induces the positive-definite metric
$h_L:=h_E|_{L\times L}$ on $L$.  We write
$u^\flat=h_L(u,\cdot)$ for $u\in\Gamma(L)$.
The generalized metric also supplies a useful \emph{algebraic},
not automatically differential, bridge between the two Clifford
structures.  Define
\begin{equation}\label{eq:j-metric-embedding}
 j_{\mathcal G}(u):=\frac1{\sqrt2}(u+\mathcal G u)
 \quad (u\in L).
\end{equation}
Since $L$ is isotropic and $\mathcal G$ is orthogonal for the
natural pairing, $j_{\mathcal G}$ is a fibrewise isometry
$\bigl(L,h_L\bigr)\xrightarrow{\sim}
\bigl(V_+,\langle\,,\,\rangle|_{V_+}\bigr)$:
\begin{equation}\label{eq:j-isometry}
\langle j_{\mathcal G}(u),j_{\mathcal G}(v)\rangle
 =h_L(u,v).
\end{equation}
Consequently $c_E(j_{\mathcal G}(u))$ satisfies the
$\mathrm{Cl}(L,h_L)$ anticommutation rule on any ambient
$\mathrm{Cl}(E)$ module, whereas $c_E(u)$ with $u\in L$ has
vanishing anticommutators. The restriction of an ambient Clifford
module to $V_+$ need not be an irreducible intrinsic spinor module,
and $j_{\mathcal G}$ alone does not relate the ambient and intrinsic
spin connections.
If $\mathcal K\in\Gamma(L)$ is also an ambient generalized Killing
section, the bracket action is algebraically compatible with
$j_{\mathcal G}$:
$\mathbb L^H_{\mathcal K}j_{\mathcal G}(u)
=j_{\mathcal G}([\mathcal K,u]_{L,H})$.
Section~\ref{sec:bridge} uses this to compare the spinorial actions
on the positive spin module.
For intrinsic computations on the Dirac structure, one can use the
metric, torsion-free Lie algebroid connection $\nabla^L$ defined by
the Koszul formula \cite{Boucetta}
\[
\begin{aligned}
2h_L(\nabla^L_u v,w)={}&\rho_L(u)h_L(v,w)
+\rho_L(v)h_L(u,w)-\rho_L(w)h_L(u,v)\\
&-h_L(u,[v,w]_{L,H})+h_L(v,[w,u]_{L,H})
+h_L(w,[u,v]_{L,H}).
\end{aligned}
\]
In particular,
$\nabla^L_u v-\nabla^L_v u=[u,v]_{L,H}$.
A generalized connection $\mathbb D$ on $E$ restricts to a
connection on $L$ if
$\mathbb D_u\Gamma(L)\subset\Gamma(L)$ for $u\in\Gamma(L)$.
Alternatively, the $h_E$-orthogonal projection $P_L$ defines the
metric $L$-connection $P_L\mathbb D_u v$ when
$\mathbb D h_E=0$, but this projected connection need not be
torsion-free relative to $[\,,\,]_{L,H}$, and it need not coincide
with $\nabla^L$.

For any chosen $L$-connection $\nabla$ on $L$, its curvature is
\[
R^\nabla(u,v)w=\nabla_u\nabla_vw-
\nabla_v\nabla_uw-\nabla_{[u,v]_{L,H}}w.
\]
For the intrinsic metric connection $\nabla=\nabla^L$, this gives
\begin{equation}\label{eq:intrinsic-curvature}
R^L(u,v)w=\nabla^L_u\nabla^L_vw-
\nabla^L_v\nabla^L_uw-\nabla^L_{[u,v]_{L,H}}w.
\end{equation}
The Jacobi identity and Lie algebroid anchor rule ensure
$C^\infty(M)$-multilinearity.  Unlike the natural pairing of
$E$, the intrinsic metric $h_L$ allows a nontrivial curvature
four-tensor $h_L(R^L(u,v)w,z)$.

To compare this operator with an ambient generalized connection,
use a \emph{matching} $H$-twisted Dorfman bracket and define the
raw ambient operator
\begin{equation}\label{eq:raw-curvature-E}
R^{(0)}_{E,H}(u,v)w
:=\mathbb D_u\mathbb D_vw-\mathbb D_v\mathbb D_uw
-\mathbb D_{[u,v]_{D,H}}w.
\end{equation}
On the full Courant algebroid this raw expression need not be
$C^\infty(M)$-linear in all its inputs.  The tensorial
Courant-algebroid curvature therefore requires a separate
construction, with additional correction terms
\cite{HohmZwiebach,JurcoVysoky}.

There is nevertheless a precise conditional comparison.
Let $P_L:E\to L$ be the $h_E$-orthogonal projector and set
$\nabla^{\mathrm{pr}}_u v=P_L\mathbb D_u v$ for $u,v\in\Gamma(L)$.
The connection $\nabla^{\mathrm{pr}}$ is metric if
$\mathbb D h_E=0$.  Writing
$\mathrm{II}(u,v)=(1-P_L)\mathbb D_u v$, direct expansion gives
\begin{equation}\label{eq:projected-curvature}
R^{\mathrm{pr}}(u,v)w
=P_LR^{(0)}_{E,H}(u,v)w
-P_L\mathbb D_u\mathrm{II}(v,w)
+P_L\mathbb D_v\mathrm{II}(u,w).
\end{equation}
If $\mathbb D_u\Gamma(L)\subset\Gamma(L)$ for
$u\in\Gamma(L)$, then $\mathrm{II}=0$ on such directions and
$R^{\mathrm{pr}}=R^{(0)}_{E,H}|_{L}$ as $L$-valued operators.
It coincides with $R^L$ when, in addition,
$\nabla^{\mathrm{pr}}=\nabla^L$.

A simple example shows that the natural pairing does not
recover the intrinsic curvature four-tensor when all four
arguments lie in $L$.  Take $H=B=0$ and $L=TM$ on a
Riemannian manifold $(M,g)$ with nonzero Riemann curvature,
for example a round two-sphere.  Then
$h_L=\frac12 g$ and $\nabla^L=\nabla^g$.
The generalized connection
\[
\mathbb D_{X+\xi}(Y+\eta)
=\nabla^g_XY+\nabla^g_X\eta
\]
is compatible with both the natural pairing and the generalized
metric, is generalized torsion-free for the untwisted bracket,
and preserves $TM$ in all directions.  Consequently,
$R^L(X,Y)Z=R^g(X,Y)Z$ is generally nonzero, whereas
\begin{equation}\label{eq:isotropic-curvature-pairing}
\langle R^{(0)}_{E,0}(X,Y)Z,W\rangle=0,
\qquad X,Y,Z,W\in\Gamma(TM),
\end{equation}
because $TM$ is isotropic.  The additional derivative-pairing
terms of standard ambient generalized-curvature tensorializations
also vanish for this example with all four arguments in $TM$.
Therefore the ambient four-tensor with four $L$ arguments is
\emph{not} the intrinsic $L$ curvature four-tensor in general.

The integrability identities below therefore use $R^L$ and the
intrinsic torsion-free connection, rather than a restriction of
an ambient curvature four-tensor.

Let $\{\mathcal X_a\}_{a=1}^n$ be a local $h_L$-orthonormal
frame of $L$, with dual coframe $\{\mathcal E^a\}_{a=1}^n$.
Lowercase indices refer to the rank-$n$ Dirac structure and
capital indices $A,B,\ldots$ to the rank-$2n$ bundle $E$.
Intrinsic $p$-forms are sections of
$\Omega^p(L):=\Gamma(\Lambda^pL^*)$.
The Lie algebroid differential is defined by
\[
\begin{aligned}
(d_L\omega)(u_0,\ldots,u_p)
={}&\sum_{i=0}^p(-1)^i\rho_L(u_i)
 \omega(u_0,\ldots,\widehat{u_i},\ldots,u_p)\\
&+\sum_{0\leq i<j\leq p}(-1)^{i+j}
\omega([u_i,u_j]_{L,H},u_0,\ldots,
\widehat{u_i},\ldots,\widehat{u_j},\ldots,u_p),
\end{aligned}
\]
so that $d_L^2=0$.  With the metric, torsion-free connection
$\nabla^L$ it can equivalently be written as
\[
d_L\omega=\mathcal E^a\wedge\nabla^L_{\mathcal X_a}\omega,
\qquad
\delta_L\omega=-i_{\mathcal X_a}\nabla^L_{\mathcal X_a}\omega.
\]
For oriented $L$, the Hodge identity
$\delta_L|_{\Omega^p(L)}=(-1)^{n(p+1)+1}*_L d_L *_L$
follows from metricity and torsion-freeness and gives $\delta_L^2=0$.
Here $\delta_L$ denotes the covariant-trace coderivative;
its identification with an $L^2$-formal adjoint requires
appropriate density and divergence conventions.
For a connection with torsion,
$\mathcal E^a\wedge\nabla_{\mathcal X_a}\omega$
need not equal $d_L\omega$ or square to zero.

The intrinsic Killing-vector equation for
$\mathcal K\in\Gamma(L)$, with
$\widetilde{\mathcal K}:=\mathcal K^\flat$, is
\begin{equation}
\nabla^L_{\mathcal{X}_a}\widetilde{\mathcal{K}}=
\frac{1}{2}i_{\mathcal{X}_a}d_L\widetilde{\mathcal{K}}.
\end{equation}
Similarly, its conformal Killing counterpart is
\begin{equation}
\nabla^L_{\mathcal{X}_a}\widetilde{\mathcal{K}}=
\frac{1}{2}i_{\mathcal{X}_a}d_L\widetilde{\mathcal{K}}
-\frac{1}{n}\mathcal{E}_a\wedge\delta_L\widetilde{\mathcal{K}}.
\end{equation}
In particular, $\delta_L\widetilde{\mathcal K}=0$
reduces the conformal Killing condition to the Killing condition.
\subsection{Intrinsic Killing--Yano and conformal Killing--Yano forms}

All formulas in this subsection are intrinsic to the metric Lie
algebroid $(L,[\,,\,]_{L,H},\rho_L,h_L)$ and its torsion-free
Levi--Civita $L$-connection $\nabla^L$. The identities below are
stated for $1\leq p\leq n-1$.

A Killing--Yano (KY) $p$-form $\omega\in\Omega^p(L)$ satisfies
\begin{equation}\label{eq:KY-intrinsic}
\nabla^L_{\mathcal X_a}\omega
=\frac1{p+1}i_{\mathcal X_a}d_L\omega.
\end{equation}
Contracting this equation gives $\delta_L\omega=0$.
The conformal Killing--Yano (CKY) equation, in the standard
convention of \cite{Semmelmann}, is
\begin{equation}\label{eq:CKY-intrinsic}
\nabla^L_{\mathcal X_a}\omega
=\frac1{p+1}i_{\mathcal X_a}d_L\omega
-\frac1{n-p+1}\mathcal E_a\wedge\delta_L\omega.
\end{equation}
The closed CKY (CCKY) subclass obeys $d_L\omega=0$.
Here and below, the curvature $R^L(u,v)$ acts on forms by the
connection induced from its action on $L$: explicitly,
\begin{equation}\label{eq:curvature-on-forms}
(R^L(u,v)\omega)(w_1,\ldots,w_p)
=-\sum_{j=1}^{p}\omega(w_1,\ldots,R^L(u,v)w_j,\ldots,w_p).
\end{equation}
This sign convention is fixed throughout the integrability
computations below.

\subsection{Integrability identities}

\begin{proposition}\label{prop:integrability}
Let $\omega\in\Omega^p(L)$, $1\leq p\leq n-1$, be a CKY form, and let
$\{\mathcal X_a\}_{a=1}^{n}$ be a local $h_L$-orthonormal frame with dual
coframe $\{\mathcal E^a\}$. Then
\begin{equation}\label{eq:CKY-d-integrability}
\nabla^L_{\mathcal X_b}d_L\omega
=-\frac{p+1}{p}\,\mathcal E^a\wedge
 R^L(\mathcal X_a,\mathcal X_b)\omega
+\frac{p+1}{p(n-p+1)}\,
 \mathcal E_b\wedge d_L\delta_L\omega.
\end{equation}
and
\begin{equation}\label{eq:CKY-delta-integrability}
\nabla^L_{\mathcal X_b}\delta_L\omega
=\frac{n-p+1}{n-p}\,
 i_{\mathcal X_a}R^L(\mathcal X_a,\mathcal X_b)\omega
-\frac{n-p+1}{(n-p)(p+1)}\,
 i_{\mathcal X_b}\delta_Ld_L\omega.
\end{equation}
\end{proposition}

\begin{proof}
We do not use a normal $L$-frame: when the anchor of a Lie algebroid
has a kernel, all connection coefficients need not vanish
simultaneously at a point. Instead, take the covariant Hessian
$\nabla^{L,2}_{u,v}\omega
:=\nabla^L_u\nabla^L_v\omega
-\nabla^L_{\nabla^L_u v}\omega$.  Torsion-freeness
gives the invariant commutator
$\nabla^{L,2}_{u,v}\omega
-\nabla^{L,2}_{v,u}\omega=R^L(u,v)\omega$.
Differentiating \eqref{eq:CKY-intrinsic} tensorially
and antisymmetrizing gives
\begin{equation}\label{eq:CKY-curvature-commutator}
\begin{aligned}
R^L(\mathcal X_a,\mathcal X_b)\omega
={}&\frac1{p+1}\left(
  i_{\mathcal X_b}\nabla^L_{\mathcal X_a}d_L\omega
 -i_{\mathcal X_a}\nabla^L_{\mathcal X_b}d_L\omega\right)\\
&-\frac1{n-p+1}\left(
  \mathcal E_b\wedge\nabla^L_{\mathcal X_a}\delta_L\omega
 -\mathcal E_a\wedge\nabla^L_{\mathcal X_b}\delta_L\omega
\right).
\end{aligned}
\end{equation}
Wedge \eqref{eq:CKY-curvature-commutator} on the left by
$\mathcal E^a$ and sum over $a$. Using
\[
\mathcal E^a\wedge i_{\mathcal X_a}\alpha
  =(\deg\alpha)\alpha,\qquad d_L^2=0,
\]
and $\mathcal E^a\wedge\nabla^L_{\mathcal X_a}\alpha
=d_L\alpha$, one obtains \eqref{eq:CKY-d-integrability}.
Contracting \eqref{eq:CKY-curvature-commutator} with
$i_{\mathcal X_a}$ and summing over $a$, the identities
$\delta_L=-i_{\mathcal X_a}\nabla^L_{\mathcal X_a}$,
$\delta_L^2=0$, and
$i_{\mathcal X_a}(\mathcal E^a\wedge\alpha)
=(n-\deg\alpha)\alpha$ lead to
\eqref{eq:CKY-delta-integrability}; the division by $n-p$ uses $p<n$.
\end{proof}

For KY forms, $\delta_L\omega=0$ and \eqref{eq:CKY-d-integrability} gives
\begin{equation}\label{eq:KY-integrability}
\nabla^L_{\mathcal X_b}d_L\omega
=-\frac{p+1}{p}\,\mathcal E^a\wedge
R^L(\mathcal X_a,\mathcal X_b)\omega.
\end{equation}
For $p=1$, the Killing one-form
$\widetilde{\mathcal K}=\mathcal K^\flat$ obeys
\begin{equation}\label{eq:Killing-1form-integrability}
\nabla^L_{\mathcal X_b}d_L\widetilde{\mathcal K}
=-2\mathcal E^a\wedge
 R^L(\mathcal X_a,\mathcal X_b)\widetilde{\mathcal K}.
\end{equation}
In the CCKY case $d_L\omega=0$, \eqref{eq:CKY-delta-integrability}
reduces to
\begin{equation}\label{eq:CCKY-integrability}
\nabla^L_{\mathcal X_b}\delta_L\omega
=\frac{n-p+1}{n-p}\,
 i_{\mathcal X_a}R^L(\mathcal X_a,\mathcal X_b)\omega.
\end{equation}
For top-degree forms the factor $n-p$ vanishes, and
\eqref{eq:CKY-delta-integrability} does not determine
$\nabla^L\delta_L\omega$.

\subsection{Special KY and CCKY forms}

Special KY and CCKY forms satisfy additional first-order equations
for $d_L\omega$ and $\delta_L\omega$. A KY $p$-form is
\emph{special} with constant $c$ if
\begin{equation}\label{eq:special-KY}
\nabla^L_{\mathcal X_b}d_L\omega
=-c(p+1)\,\mathcal E_b\wedge\omega.
\end{equation}
Likewise, a closed CKY $p$-form is \emph{special} if
\begin{equation}\label{eq:special-CCKY}
\nabla^L_{\mathcal X_b}\delta_L\omega
=c(n-p+1)\,i_{\mathcal X_b}\omega.
\end{equation}
The $p$-dependent normalization of $c$ is chosen so that a
space-form test has the same $c$ in the KY and CCKY sectors.
By \eqref{eq:KY-integrability} and
\eqref{eq:CCKY-integrability}, the corresponding necessary
curvature conditions are
\begin{equation}\label{eq:special-curvature-conditions}
\begin{aligned}
\mathcal E^a\wedge R^L(\mathcal X_a,\mathcal X_b)\omega
&=pc\,\mathcal E_b\wedge\omega
&&\text{(special KY)},\\
i_{\mathcal X_a}R^L(\mathcal X_a,\mathcal X_b)\omega
&=(n-p)c\,i_{\mathcal X_b}\omega
&&\text{(special CCKY)}.
\end{aligned}
\end{equation}
These conditions are necessary; they do not by themselves
guarantee the existence of special forms.

As a consistency check, suppose the intrinsic
connection has constant sectional curvature $k$ with
\begin{equation}\label{eq:spaceform-curvature}
R^L(u,v)w=k\bigl(h_L(v,w)u-h_L(u,w)v\bigr).
\end{equation}
Its induced action on $p$-forms obeys
\begin{equation}\label{eq:spaceform-form-contractions}
\mathcal E^a\wedge R^L(\mathcal X_a,\mathcal X_b)\omega
=pk\,\mathcal E_b\wedge\omega,\qquad
 i_{\mathcal X_a}R^L(\mathcal X_a,\mathcal X_b)\omega
=(n-p)k\,i_{\mathcal X_b}\omega.
\end{equation}
Therefore every KY/CCKY form of the appropriate degree
on this constant-curvature Lie algebroid obeys the special
equations with $c=k$.

In Section~\ref{sec:spinor-symmetry} we use the positive Clifford
convention $\{c_L(u),c_L(v)\}=2h_L(u,v)$. Killing-spinor
integrability then gives $k=-4\lambda^2$ in a space form.
Thus $c=-4\lambda^2$ is the same in both special-form equations.
The negative-square Riemannian convention instead gives
$k=4\lambda^2$ with its correspondingly rescaled Killing number.
Section~\ref{sec:bilinears} derives $c=-4\lambda^2$ without
a space-form assumption, for the specified Hermitian bilinears.

\subsection{Flux in the intrinsic connection}
The closed three-form $H$ enters the intrinsic geometry through the
twisted bracket $[\,,\,]_{L,H}$ and therefore through its unique
metric, torsion-free Lie-algebroid connection. This prescription yields
well-defined curvature and exterior-trace identities even for
noninvertible anchors. A distinct but related extension is to choose
an additional skew-torsion connection; the resulting differential
operators and integrability conditions require their own torsion terms
\cite{HouriKubiznakWarnickYasui,Chrysikos}.

\section{Intrinsic Spinors and Symmetry Operators}
\label{sec:spinor-symmetry}

\subsection{The intrinsic spinor module: assumptions and normalization}

The preceding KY/CKY integrability equations concern
$\Omega^\bullet(L)$ and the metric connection $\nabla^L$.
To act on their solutions by Clifford multiplication, we now
\emph{assume} that the oriented, rank-$n$ metric Lie algebroid
$(L,h_L)$ carries a spin structure (or work locally on its spin representation) and let $S_L$ be an intrinsic spinor
bundle with Clifford action $c_L$ and Levi--Civita lift
$\nabla^S$.  We impose
\begin{equation}\label{eq:intrinsic-clifford}
\{c_L(u),c_L(v)\}=2h_L(u,v),\qquad
[\nabla^S_u,c_L(v)]=c_L(\nabla^L_u v).
\end{equation}
In an orthonormal frame write $\gamma_a=c_L(\mathcal X_a)$.
We use the standard ordered-product map from differential forms
to the Clifford algebra; in particular,
$c_L(\mathcal E^a\wedge\mathcal E^b)=\gamma_a\gamma_b$
for $a\ne b$.  The identities
\begin{equation}\label{eq:intrinsic-clifford-identities}
\begin{aligned}
\gamma_b c_L(\omega)&=c_L(\mathcal E_b\wedge\omega)
                         +c_L(i_{\mathcal X_b}\omega),\\
c_L(\omega)\gamma_b&=(-1)^p
\bigl(c_L(\mathcal E_b\wedge\omega)-c_L(i_{\mathcal X_b}\omega)\bigr),
&&\omega\in\Omega^p(L),\\
c_L(i_{\mathcal X_a}\omega)\gamma_a
&=(-1)^{p-1}p\,c_L(\omega)
\end{aligned}
\end{equation}
fix the numerical factors throughout. The isometry
$j_{\mathcal G}:L\to V_+$ transports the chosen spin structure
to $V_+$; let $S(V_+)$ be its associated spin module. These
are ordinary complex spin bundles with their principal spin lifts,
without an auxiliary line twist. In odd rank, fix one extension of
the spin representation to the complex Clifford algebra.

An intrinsic Killing spinor with Killing number $\lambda\in\mathbb C$
satisfies
\begin{equation}\label{eq:intrinsic-Killing-spinor}
\nabla^S_{\mathcal X_a}\psi=\lambda\gamma_a\psi.
\end{equation}
Let $\not\!D_L:=\gamma_a\nabla^S_{\mathcal X_a}$.  The
intrinsic twistor equation, under our positive Clifford convention,
is
\begin{equation}\label{eq:intrinsic-twistor-spinor}
\nabla^S_{\mathcal X_a}\psi
=\frac1n\gamma_a\not\!D_L\psi.
\end{equation}
Accordingly, a Killing spinor satisfies
$\not\!D_L\psi=n\lambda\psi$ and is a twistor spinor,
but a general twistor spinor need not be harmonic
($\not\!D_L\psi=0$).

\subsection{The sign of the special-form constant}

The Clifford sign has an immediate consequence.  Applying the
$L$-curvature to \eqref{eq:intrinsic-Killing-spinor} gives, for
$a\ne b$,
\begin{equation}\label{eq:spinor-integrability}
R^S(\mathcal X_a,\mathcal X_b)\psi
=\lambda^2(\gamma_b\gamma_a-\gamma_a\gamma_b)\psi
=-2\lambda^2\gamma_a\gamma_b\psi.
\end{equation}
For the space-form curvature convention
\eqref{eq:spaceform-curvature}, the spin representation of
curvature obeys
$R^S(\mathcal X_a,\mathcal X_b)=
\tfrac{k}{2}\gamma_a\gamma_b$ for $a\ne b$, as can
be checked from its commutator with $c_L(v)$.
Consequently a nonzero Killing spinor requires
\begin{equation}\label{eq:spaceform-k-lambda}
k=-4\lambda^2.
\end{equation}
Since $c=k$ for special forms on a space form, both sectors use
$c=-4\lambda^2$. For a form chosen independently of spinors this
remains an additional hypothesis outside constant curvature.
For the bilinears in Section~\ref{sec:bilinears}, it follows directly
from their differential identities.

\subsection{Killing sections and the intrinsic Kosmann derivative}

For an intrinsic Killing section $\mathcal K\in\Gamma(L)$,
$\mathcal L^{L}_{\mathcal K}h_L=0$ and its metric-dual
one-form $\widetilde{\mathcal K}=\mathcal K^\flat$
is KY of degree one.  The intrinsic Kosmann derivative is
\begin{equation}\label{eq:Kosmann-intrinsic}
\mathscr L^{S}_{\mathcal K}\psi
=\nabla^S_{\mathcal K}\psi
+\frac14 c_L(d_L\widetilde{\mathcal K})\psi.
\end{equation}
The factor $1/4$ corresponds to the ordered-product normalization
of $c_L$ in \eqref{eq:intrinsic-clifford-identities}, applied to the
two-form $d_L\widetilde{\mathcal K}$.
Let $A^L_{\mathcal K}u=[\mathcal K,u]_{L,H}-\nabla^L_{\mathcal K}u
=-\nabla^L_u\mathcal K$.  The spin lift
$-\tfrac14(A^L_{\mathcal K})_{ab}\gamma_a\gamma_b$
precisely yields the algebraic term in
\eqref{eq:Kosmann-intrinsic}, and therefore
\begin{equation}\label{eq:Kosmann-intrinsic-clifford}
[\mathscr L^S_{\mathcal K},c_L(u)]
=c_L([\mathcal K,u]_{L,H}).
\end{equation}
The Levi--Civita $L$-connection is invariant under an intrinsic
Killing section, so its lifted connection obeys
\begin{equation}\label{eq:Kosmann-connection-equivariance}
[\mathscr L^S_{\mathcal K},\nabla^S_u]
=\nabla^S_{[\mathcal K,u]_{L,H}}.
\end{equation}
Combining \eqref{eq:Kosmann-intrinsic-clifford} and
\eqref{eq:Kosmann-connection-equivariance} with
\eqref{eq:intrinsic-Killing-spinor} shows directly that
\begin{equation}\label{eq:Kosmann-Killing-preservation}
\nabla^S_u(\mathscr L^S_{\mathcal K}\psi)
=\lambda c_L(u)\mathscr L^S_{\mathcal K}\psi.
\end{equation}
This holds for every intrinsic Killing section, without a curvature
assumption. The following theorem identifies its positive-spinor
counterpart for ambient generalized Killing sections.

\subsection{Comparison with the positive spinor action}\label{sec:bridge}

\begin{theorem}\label{prop:bridge}
Let $L$ be an $H$-twisted Dirac structure with metric $h_L$ and a
chosen spin structure. Transport this spin structure to $V_+$ by
$j_{\mathcal G}$, and denote the induced spinor isomorphism by
$\Phi:S_L\to S(V_+)$. Suppose $\mathcal K\in\Gamma(L)$
satisfies $\mathbb L^H_{\mathcal K}\mathcal G=0$.
For any pairing- and generalized-metric-compatible generalized
connection $\mathbb D$, its induced spin connection on $S(V_+)$
and the infinitesimal spin representation $\sigma_+$ define
\begin{equation}\label{eq:positive-lift}
\mathbb L^{\mathrm{spin},+}_{\mathcal K}
=\mathbb D^{S(V_+)}_{\mathcal K}
+\sigma_+(\mathcal A_{\mathcal K}|_{V_+}).
\end{equation}
Then $\mathcal K$ is intrinsically Killing and
\begin{equation}\label{eq:positive-intertwining}
\Phi^{-1}\mathbb L^{\mathrm{spin},+}_{\mathcal K}\Phi
=\mathscr L^S_{\mathcal K}.
\end{equation}
In particular this action is independent of $\mathbb D$.
For a torsion-free compatible $\mathbb D$ it is the positive-spinor
KD action with the conventions of Section~\ref{sec:ambient}.
\end{theorem}

\begin{proof}
Involutivity of $L$ and the generalized Killing condition give
\begin{equation}\label{eq:j-bracket-equivariance}
\mathbb L^H_{\mathcal K}j_{\mathcal G}(u)
=j_{\mathcal G}([\mathcal K,u]_{L,H}).
\end{equation}
Preservation of the ambient pairing and $\mathcal G$ therefore
implies $\mathcal L^L_{\mathcal K}h_L=0$.
Pull back the $V_+$ connection to the metric $L$-connection
$\nabla^+_u v=j_{\mathcal G}^{-1}\mathbb D_u j_{\mathcal G}(v)$.
Its difference from $\nabla^L$ is a skew endomorphism:
$C_u=\nabla^+_u-\nabla^L_u\in\mathfrak{so}(L,h_L)$.
Equation~\eqref{eq:j-bracket-equivariance} gives
\[
j_{\mathcal G}^{-1}(\mathcal A_{\mathcal K}|_{V_+})j_{\mathcal G}
=A^L_{\mathcal K}-C_{\mathcal K}.
\]
For $A\in\mathfrak{so}(L,h_L)$ write
$\sigma(A)=-\tfrac14h_L(A\mathcal X_a,\mathcal X_b)
\gamma_a\gamma_b$. Naturality of the principal spin lift yields
\begin{equation}\label{eq:connection-cancellation}
\begin{aligned}
\Phi^{-1}\mathbb L^{\mathrm{spin},+}_{\mathcal K}\Phi
&=\nabla^S_{\mathcal K}+\sigma(C_{\mathcal K})
  +\sigma(A^L_{\mathcal K}-C_{\mathcal K})\\
&=\nabla^S_{\mathcal K}+\sigma(A^L_{\mathcal K})
 =\mathscr L^S_{\mathcal K}.
\end{aligned}
\end{equation}
The cancellation uses neither a torsion hypothesis nor the equality
$\nabla^+=\nabla^L$; generalized torsion-freeness enters only when
the generator is expressed through the covariant KD projection.
\end{proof}

The theorem compares infinitesimal actions on $S(V_+)$. A realization
that also intertwines the spin connections can be chosen as follows.
Since $E=L\oplus\mathcal G L$, prescribe along $u\in\Gamma(L)$
\begin{equation}\label{eq:adapted-connection}
\mathbb D_u(v+\mathcal G w)
=\nabla^L_u v+\mathcal G\nabla^L_u w,\qquad v,w\in\Gamma(L).
\end{equation}
Isotropy and $h_L(v,w)=\langle v,\mathcal G w\rangle$ show
that this preserves the pairing and $\mathcal G$. Extend it to
directions in $\mathcal G L$ using any compatible reference
connection; the direct sum gives a generalized connection on $E$.
Then $\nabla^+=\nabla^L$ and $\Phi$ intertwines their induced
spin connections along $L$. Thus the intrinsic Killing equation
transports to
\[
\mathbb D^{S(V_+)}_u\Phi\psi
=\lambda c_+(j_{\mathcal G}u)\Phi\psi,\qquad u\in\Gamma(L).
\]

\begin{remark}\label{rem:adapted}
The adapted connection \eqref{eq:adapted-connection} need not be
generalized torsion-free. Moreover, an intrinsic Killing section of $L$
need not satisfy the ambient condition
$\mathbb L^H_{\mathcal K}\mathcal G=0$ of Theorem~\ref{prop:bridge};
an explicit example is given in Section~\ref{sec:nontrivial-flux}.
\end{remark}

\subsection{Hermitian Killing-spinor bilinears}\label{sec:bilinears}

Fix the Spin-invariant Hermitian product $(\,,\,)$ on $S_L$,
conjugate-linear in its first argument, with $c_L(u)$ self-adjoint
for real $u$. The Levi--Civita spin connection preserves this
product. Following the differential-form bilinear construction
of \cite{AcikErtem}, specify the Clifford order explicitly by
\begin{equation}\label{eq:bilinear-definition}
\omega_p(\psi,\phi)
=\frac1{p!}(\psi,\gamma_{[a_1}\cdots\gamma_{a_p]}\phi)
\mathcal E^{a_1}\wedge\cdots\wedge\mathcal E^{a_p},
\end{equation}
where brackets denote normalized antisymmetrization. These are
complex-valued forms; they are sesquilinear in the spinors.

\begin{proposition}\label{prop:bilinears}
Let $\psi$ and $\phi$ satisfy the intrinsic Killing equation
with the same constant $\lambda\in i\mathbb R$.
For odd $p$, $\omega_p$ is KY, and for even $p$, it is CCKY.
In the degree ranges used in the symmetry theorems, both are
special with $c=-4\lambda^2$, without a constant-curvature
assumption.
\end{proposition}

\begin{proof}
Take the covariant derivative of \eqref{eq:bilinear-definition}
as a tensor; no normal $L$-frame is required. For an antisymmetric
multi-index $I$, self-adjointness of $\gamma_b$ gives
\[
(\nabla^L_{\mathcal X_b}\omega_p)_I
=\bar\lambda(\psi,\gamma_b\gamma_I\phi)
 +\lambda(\psi,\gamma_I\gamma_b\phi).
\]
Since $\bar\lambda=-\lambda$, the right side is the Clifford
commutator. Expanding it yields
\begin{equation}\label{eq:bilinear-recursion}
\nabla^L_{\mathcal X_b}\omega_p=
\begin{cases}
-2\lambda i_{\mathcal X_b}\omega_{p+1},&p\ \text{odd},\\
-2\lambda\mathcal E_b\wedge\omega_{p-1},&p\ \text{even}.
\end{cases}
\end{equation}
Set $\omega_{-1}=\omega_{n+1}=0$ at the endpoints.
Exterior and contraction traces give
\begin{equation}\label{eq:bilinear-traces}
\begin{aligned}
d_L\omega_p&=-2\lambda(p+1)\omega_{p+1},&
\delta_L\omega_p&=0 &&(p\ \text{odd}),\\
d_L\omega_p&=0,&
\delta_L\omega_p&=2\lambda(n-p+1)\omega_{p-1}
&&(p\ \text{even}).
\end{aligned}
\end{equation}
Applying \eqref{eq:bilinear-recursion} once more gives
$\nabla^L_{\mathcal X_b}d_L\omega_p
=4\lambda^2(p+1)\mathcal E_b\wedge\omega_p$ for odd $p$,
and $\nabla^L_{\mathcal X_b}\delta_L\omega_p
=-4\lambda^2(n-p+1)i_{\mathcal X_b}\omega_p$ for even $p$.
These are precisely the two special equations with
$c=-4\lambda^2$.
\end{proof}

\begin{remark}\label{rem:bilinears}
The Clifford order in \eqref{eq:bilinear-definition} differs from
the reversed order used in some Dirac-current conventions by
$(-1)^{p(p-1)/2}$; this changes signs in the adjacent-degree
recursions but not the special constant. The proof uses no Fierz
completeness and therefore also applies to an irreducible odd-rank
spin module. For real nonzero $\lambda$ the Hermitian pairing
produces an anticommutator instead, and the parity assignment of
Proposition~\ref{prop:bilinears} does not carry over unchanged. For
$\lambda=0$ all these bilinears are parallel.
\end{remark}

\subsection{Odd special KY Killing-spinor symmetries}

The operator below and its even-CCKY counterpart have the same
coefficients as \cite[Theorem~3, Eq.~(89)]{AcikErtem}. We give the
intrinsic Lie-algebroid proof.

\begin{theorem}\label{thm:KY-operator}
Let $p$ be odd, $1\le p\le n-1$, and let
$\omega\in\Omega^p(L)$ satisfy the KY equation
\eqref{eq:KY-intrinsic} and the special condition
\eqref{eq:special-KY}, with
\begin{equation}\label{eq:KY-special-c}
c=-4\lambda^2.
\end{equation}
Then the operator on $S_L$ defined by
\begin{equation}\label{eq:operator-KY-proved}
\mathscr L^{\mathrm{KY}}_{\omega}\psi
=c_L(i_{\mathcal X_a}\omega)\nabla^S_{\mathcal X_a}\psi
+\frac{p}{2(p+1)}c_L(d_L\omega)\psi.
\end{equation}
maps intrinsic Killing spinors with Killing number $\lambda$ to
intrinsic Killing spinors with the same Killing number.
\end{theorem}

\begin{proof}
For a Killing spinor, the last identity in
\eqref{eq:intrinsic-clifford-identities} gives
\begin{equation}\label{eq:KY-operator-on-KS}
\mathscr L^{\mathrm{KY}}_{\omega}\psi
=p\lambda c_L(\omega)\psi
+\frac{p}{2(p+1)}c_L(d_L\omega)\psi.
\end{equation}
Differentiating, using the KY equation and the special condition,
and subtracting the Killing-spinor equation for the output,
one obtains the following unreduced residual:
\begin{equation}\label{eq:KY-residual-full}
\begin{aligned}
\nabla^S_{\mathcal X_b}(\mathscr L^{\mathrm{KY}}_{\omega}\psi)
-\lambda\gamma_b\mathscr L^{\mathrm{KY}}_{\omega}\psi
={}&\lambda\frac{p}{p+1}c_L(i_{\mathcal X_b}d_L\omega)\psi
+\lambda^2 p[c_L(\omega),\gamma_b]\psi\\
&-\frac{cp}{2}c_L(\mathcal E_b\wedge\omega)\psi
+\lambda\frac{p}{2(p+1)}[c_L(d_L\omega),\gamma_b]\psi.
\end{aligned}
\end{equation}
Since $p$ is odd, $d_L\omega$ is even and
\[
[c_L(\omega),\gamma_b]=-2c_L(\mathcal E_b\wedge\omega),
\quad[c_L(d_L\omega),\gamma_b]=-2c_L(i_{\mathcal X_b}d_L\omega).
\]
Thus the $i_{\mathcal X_b}d_L\omega$ terms cancel and
\begin{equation}\label{eq:KY-residual-final}
\nabla^S_{\mathcal X_b}(\mathscr L^{\mathrm{KY}}_{\omega}\psi)
-\lambda\gamma_b\mathscr L^{\mathrm{KY}}_{\omega}\psi
=-\frac p2(c+4\lambda^2)
 c_L(\mathcal E_b\wedge\omega)\psi=0.
\end{equation}
\end{proof}

For $p=1$, \eqref{eq:operator-KY-proved} reduces exactly to
\eqref{eq:Kosmann-intrinsic}; in that case Killing spinors are
preserved for every intrinsic Killing section, without an additional
special-form equation.

\subsection{Even special CCKY Killing-spinor symmetries}

\begin{theorem}\label{thm:CCKY-operator}
Let $p$ be even, $2\le p\le n-1$, and let $\omega\in\Omega^p(L)$
be a closed CKY form, that is, a solution of \eqref{eq:CKY-intrinsic}
with $d_L\omega=0$, satisfying the special condition
\eqref{eq:special-CCKY} with $c=-4\lambda^2$. Then the operator
\begin{equation}\label{eq:operator-CCKY-proved}
\mathscr L^{\mathrm{CCKY}}_{\omega}\psi
=c_L(i_{\mathcal X_a}\omega)\nabla^S_{\mathcal X_a}\psi
+\frac{p}{2(n-p+1)}c_L(\delta_L\omega)\psi.
\end{equation}
maps intrinsic Killing spinors with Killing number $\lambda$ to
intrinsic Killing spinors with the same Killing number.
\end{theorem}

\begin{proof}
For a Killing spinor the Clifford contraction identity yields
\begin{equation}\label{eq:CCKY-operator-on-KS}
\mathscr L^{\mathrm{CCKY}}_{\omega}\psi
=-p\lambda c_L(\omega)\psi
+\frac{p}{2(n-p+1)}c_L(\delta_L\omega)\psi.
\end{equation}
Differentiating and using the CCKY equation gives the
intermediate residual
\begin{equation}\label{eq:CCKY-residual-full}
\begin{aligned}
\nabla^S_{\mathcal X_b}(\mathscr L^{\mathrm{CCKY}}_{\omega}\psi)
-\lambda\gamma_b\mathscr L^{\mathrm{CCKY}}_{\omega}\psi
={}&\lambda\frac{p}{n-p+1}c_L(\mathcal E_b\wedge\delta_L\omega)\psi
-\lambda^2p[c_L(\omega),\gamma_b]\psi\\
&+\frac{cp}{2}c_L(i_{\mathcal X_b}\omega)\psi
+\lambda\frac{p}{2(n-p+1)}[c_L(\delta_L\omega),\gamma_b]\psi.
\end{aligned}
\end{equation}
For even $p$ the relevant commutators are
\[
[c_L(\omega),\gamma_b]=-2c_L(i_{\mathcal X_b}\omega),
\quad[c_L(\delta_L\omega),\gamma_b]
=-2c_L(\mathcal E_b\wedge\delta_L\omega).
\]
The wedge terms cancel, leaving
\begin{equation}\label{eq:CCKY-residual-final}
\nabla^S_{\mathcal X_b}(\mathscr L^{\mathrm{CCKY}}_{\omega}\psi)
-\lambda\gamma_b\mathscr L^{\mathrm{CCKY}}_{\omega}\psi
=\frac p2(c+4\lambda^2)c_L(i_{\mathcal X_b}\omega)\psi=0.
\end{equation}
\end{proof}

\subsection{Twistor and intrinsic Dirac equations}

The intrinsic twistor operator is
$P^L_u\psi:=\nabla^S_u\psi-\tfrac1n c_L(u)\not\!D_L\psi$.
For an intrinsic Killing section $\mathcal K$, the
metric/spin equivariance in
\eqref{eq:Kosmann-intrinsic-clifford}--
\eqref{eq:Kosmann-connection-equivariance} yields
\begin{equation}\label{eq:Dirac-Kosmann-commutator}
[\mathscr L^S_{\mathcal K},\not\!D_L]=0,
\quad
[\mathscr L^S_{\mathcal K},P^L_u]
=P^L_{[\mathcal K,u]_{L,H}}.
\end{equation}
Thus the intrinsic Kosmann derivative preserves both the
whole twistor-spinor solution space and \emph{every} eigenspace
of the intrinsic Dirac operator (including its kernel).

The higher-degree operators preserve the Killing-spinor subspace
of $\not\!D_L\psi=n\lambda\psi$. They need not commute with
$\not\!D_L$ on all spinors. For a direct counterexample, take
flat $L=T\mathbb R^3$ with its Euclidean metric,
$\lambda=0$, and the parallel special CCKY form
$\omega=dx^1\wedge dx^2$ ($c=0$). Its operator is
\[
\mathscr L^{\mathrm{CCKY}}_\omega
=\gamma_2\partial_1-\gamma_1\partial_2.
\]
For $\chi=e^{mx^1}v$, where $m\ne0$ and $\gamma_1v=v$,
\begin{equation}\label{eq:Dirac-counterexample}
\not\!D_L\chi=m\chi,\qquad
\not\!D_L(\mathscr L^{\mathrm{CCKY}}_\omega\chi)
=-m\mathscr L^{\mathrm{CCKY}}_\omega\chi\ne
m\mathscr L^{\mathrm{CCKY}}_\omega\chi.
\end{equation}
Thus even this simple member of the family does not preserve
a general fixed nonzero Dirac eigenvalue.
Full Dirac symmetry operators require their own intertwining
identities \cite{BennCharlton,CarigliaKrtousKubiznak}.
Likewise, no all-degree twistor theorem follows from Killing-spinor
preservation, and a general twistor spinor need not be massless
\cite{Ertem2}.

\subsection{Positive-spinor realization of the intrinsic operators}
Theorem~\ref{prop:bridge} identifies the degree-one action on
$S(V_+)$ for generalized Killing sections in $L$. With the
adapted connection \eqref{eq:adapted-connection}, the higher-form
operators of Theorems~\ref{thm:KY-operator} and
\ref{thm:CCKY-operator} can also be transported by $\Phi$ along
$L$, which gives a positive-spinor realization of the intrinsic
theory along the directions of $L$.

\section{Explicit $H$-twisted examples}\label{sec:example}
\subsection{A graph Dirac structure with nonzero exact twist}

Let $(M,h)$ be an oriented spin Riemannian manifold, let $B$ be
a global two-form, and set
\begin{equation}\label{eq:graph-data}
H=-dB,\qquad L=e^B TM,\qquad
\mathcal G=e^B\mathcal G_0e^{-B},\qquad
\mathcal G_0=\begin{pmatrix}0&(2h)^{-1}\\2h&0\end{pmatrix},
\end{equation}
where $e^B(X+\xi)=X+\xi+i_XB$.
The factor $2$ ensures that $h_L$ pulls back to $h$, with the
half-pairing of \eqref{bilinear}. Write $\iota_B X=X+i_XB$.
Direct expansion of the twisted Dorfman bracket gives
\begin{equation}\label{eq:graph-bracket}
[\iota_BX,\iota_BY]_{D,H}
=\iota_B[X,Y]+i_Yi_X(dB+H)=\iota_B[X,Y].
\end{equation}
Hence $\iota_B:TM\to L$ is a metric Lie algebroid isomorphism
and transports $\nabla^h$, ordinary forms and spinors to the
intrinsic objects used above.

An explicit compatible generalized connection is
\begin{equation}\label{eq:graph-connection}
\mathbb D_{e^B(X+\xi)}\bigl(e^B(Y+\eta)\bigr)
=e^B\bigl(\nabla^h_XY+\nabla^h_X\eta\bigr).
\end{equation}
The untwisted connection inside this expression has zero generalized
torsion: substitution in the torsion definition cancels the
covector-derivative terms and leaves the ordinary torsion of
$\nabla^h$. Since $e^B$ intertwines the untwisted bracket with
the $H=-dB$ bracket, \eqref{eq:graph-connection} has
$T^{(H)}_{\mathbb D}=0$. It preserves $L$ and
$\mathbb D_{\iota_BX}\iota_BY=\iota_B\nabla^h_XY$.
Moreover
$\mathbb D_{\iota_BX}j_{\mathcal G}(\iota_BY)
=j_{\mathcal G}(\iota_B\nabla^h_XY)$,
so here the spin connections, as well as the degree-one actions,
intertwine. Every Killing vector $K$ of $h$ gives an ambient
generalized Killing section $\iota_BK$.
The corresponding polyform identities fix the signs independently:
\begin{equation}\label{eq:B-spin-transform}
c_E(e^Bu)=e^{-B\wedge}c_E(u)e^{B\wedge},\qquad
d_H(e^{-B\wedge}\alpha)=e^{-B\wedge}d\alpha.
\end{equation}
For a Killing vector and $\mu=\operatorname{vol}_{2h}$,
\eqref{eq:density-correction} consequently reduces to
$\mathbb L^{\mathrm{spin}}_{\iota_BK}
=e^{-B\wedge}\mathcal L_K e^{B\wedge}$ in the density
trivialization. This polyform action remains distinct from its
single positive-spinor representation.

To make the spinors and forms explicit, take $M=SU(2)\simeq S^3$
with a global $h$-orthonormal left-invariant frame $E_a$ satisfying
$[E_a,E_b]=\varepsilon_{ab}{}^cE_c$. Its dual coframe $\theta^a$
obeys $d\theta^1=-\theta^2\wedge\theta^3$ and cyclic permutations.
The bi-invariant metric $h$ has
\begin{equation}\label{eq:S3-connection}
\nabla^h_{E_a}E_b=\tfrac12\varepsilon_{ab}{}^cE_c,
\qquad k=\tfrac14.
\end{equation}
Choose a smooth global $f$ with $E_1(f)\not\equiv0$ and put
\begin{equation}\label{eq:S3-flux}
B=f\,\theta^2\wedge\theta^3,
\qquad H=-E_1(f)\,\theta^1\wedge\theta^2\wedge\theta^3\not\equiv0.
\end{equation}
For example, in unit-quaternion coordinates
$q=q_0+q_1\mathbf i+q_2\mathbf j+q_3\mathbf k$ choose
$f=q_0$ and $E_1(q)=q\mathbf i/2$, so $E_1(f)=-q_1/2$.
This is a nonzero differential form with $[H]=0$, so the example
is globally $B$-equivalent to the untwisted one; a flux with
nonzero cohomology class is treated in
Section~\ref{sec:nontrivial-flux}.

In the global spin frame, choose the Pauli matrices
$\gamma_1,\gamma_2,\gamma_3$, with
$\gamma_1\gamma_2=i\gamma_3$. The spin connection is
\begin{equation}\label{eq:S3-spin}
\nabla^S_{E_a}=E_a-\frac{i}{4}\gamma_a.
\end{equation}
Every constant two-component spinor is therefore a global
Killing spinor with $\lambda=-i/4$, and
$-4\lambda^2=1/4=k$. Via $\iota_B$, these are intrinsic
spinors of the graph Dirac structure and, via $\Phi$, spinors
of $S(V_+)$ with the chosen connection.

The forms $\kappa=\theta^3$ and
$\omega=\theta^1\wedge\theta^2$ give a special KY one-form
and a special CCKY two-form, respectively. Directly,
\begin{equation}\label{eq:S3-forms}
d\kappa=-\theta^1\wedge\theta^2,\quad \delta\kappa=0,
\qquad d\omega=0,\quad\delta\omega=-\theta^3.
\end{equation}
Using \eqref{eq:S3-connection} verifies their defining equations
and the special conditions with $c=1/4$. On constant spinors
the two operators are explicitly
\begin{equation}\label{eq:S3-actions}
\mathscr L^S_{E_3}\psi=-\frac{i}{2}\gamma_3\psi,
\qquad \mathscr L^{\mathrm{CCKY}}_\omega\psi=-\gamma_3\psi.
\end{equation}
Both outputs are constant and nonzero for every nonzero $\psi$.
For $\psi=(1,0)^{\mathsf T}$ the bilinears are
$\omega_1(\psi,\psi)=\theta^3$ and
$\omega_2(\psi,\psi)=i\theta^1\wedge\theta^2$, verifying
the recursion and its complex phases in this example.

\subsection{Nontrivial flux and a noninvertible anchor}
\label{sec:nontrivial-flux}

Let $M=S^2\times S^1$, with $S^2\subset\mathbb R^3$ the unit
sphere, position vector $q=(q_1,q_2,q_3)$, round metric $g_{S^2}$,
and area form $\Omega_q(v,w)=q\cdot(v\times w)$.
Write $dt$ for the global angular one-form on $S^1$, normalized
by $\int_{S^1}dt=2\pi$; $t$ itself is only a local angular coordinate.
With all tensors pulled back to $M$, set
\begin{equation}\label{eq:flux-product-data}
H=-\tfrac12\Omega\wedge dt,\qquad
g=2g_{S^2}+\tfrac18dt^2,\qquad
\mathcal G=\mathcal G_{g,0},\qquad
L=TS^2\oplus\mathbb R\,dt\subset TM\oplus T^*M.
\end{equation}
The flux background is the $S^2\times S^1$ member of the Hopf
T-duality pair \cite{BouwknegtEvslinMathai},
\cite[Example~2.3]{CavalcantiGualtieri}. The computation below
determines its intrinsic spinorial properties directly, without
using T-duality.

Let $\mathbf e_a$ be the standard basis of $\mathbb R^3$, and define
the global sections
\begin{equation}\label{eq:flux-product-frame}
X_a(q)=q\times\mathbf e_a,\qquad
\mathcal X_a=X_a+\tfrac12q_a\,dt,\qquad a=1,2,3.
\end{equation}
The sign in $X_a$ fixes the bracket convention.

\begin{proposition}\label{prop:flux-product}
The data \eqref{eq:flux-product-data} define an $H$-twisted
metric Dirac structure with $[H]\ne0$ and anchor of constant rank
two. The sections $\mathcal X_a$ form a global orthonormal frame
for $h_L$ and are ambient generalized Killing sections. Choose
the spin structure trivialized by this frame and Pauli matrices
$\gamma_a$. Then
\begin{equation}\label{eq:flux-product-connections}
\nabla^L_{\mathcal X_a}\mathcal X_b
=\tfrac12\varepsilon_{ab}{}^c\mathcal X_c,\qquad
\nabla^S_{\mathcal X_a}=X_a-\tfrac i4\gamma_a,
\end{equation}
and $R^L$ has constant sectional curvature $1/4$.
The full space of intrinsic Killing spinors with $\lambda=-i/4$
is $C^\infty(S^1,\mathbb C^2)$, viewed as spinors constant in
the $S^2$ directions in this frame.
\end{proposition}

\begin{proof}
The bundle $L$ is isotropic and has rank three. The rotation fields
satisfy
\begin{equation}\label{eq:flux-product-rotations}
[X_a,X_b]=\varepsilon_{ab}{}^cX_c,\quad
X_aq_b=\varepsilon_{ab}{}^cq_c,\quad
i_{X_a}\Omega=-dq_a,\quad
\Omega(X_a,X_b)=\varepsilon_{ab}{}^cq_c.
\end{equation}
Thus the one-form part of the twisted frame bracket is
\[
\tfrac12\bigl(X_aq_b-X_bq_a-\Omega(X_a,X_b)\bigr)dt
=\tfrac12\varepsilon_{ab}{}^cq_c\,dt,
\]
so that
\begin{equation}\label{eq:flux-product-bracket}
[\mathcal X_a,\mathcal X_b]_{D,H}
=\varepsilon_{ab}{}^c\mathcal X_c.
\end{equation}
This proves involutivity. The frame is independent because the
map $a\mapsto q\times a+\tfrac12(q\cdot a)dt$ has zero kernel.
Moreover $\rho_L(\mathcal X_a)=X_a$ spans $TS^2$ and
$q_a\mathcal X_a=\tfrac12dt$ spans the kernel of the anchor.
Since $g_{S^2}(X_a,X_b)=\delta_{ab}-q_aq_b$, the half-pairing gives
\[
h_L(\mathcal X_a,\mathcal X_b)
=\tfrac12\left(2(\delta_{ab}-q_aq_b)
+8\,\tfrac14q_aq_b\right)=\delta_{ab}.
\]
The flux is closed and
$\int_M H=-4\pi^2\ne0$ in the product orientation.
Consequently it cannot be removed by a global $B$-transform;
such transforms also preserve the rank of the anchor.

Each $X_a$ is Killing for $g$ and
\[
d(\tfrac12q_a\,dt)-i_{X_a}H=0.
\]
These are precisely the generalized Killing conditions for
$\mathcal G_{g,0}$ in our twist convention. The constant metric
and \eqref{eq:flux-product-bracket} give
\eqref{eq:flux-product-connections} by the Koszul formula and
the ordered-index spin lift. Direct curvature evaluation yields
$R^L(\mathcal X_a,\mathcal X_b)\mathcal X_c
=\tfrac14(\delta_{bc}\mathcal X_a-\delta_{ac}\mathcal X_b)$.
For $\lambda=-i/4$, the Killing equation reduces to
$X_a\psi=0$ for all $a$. Since these fields span the tangent
space of every connected $S^2$ leaf, this is equivalent to
$\psi(q,t)=\chi(t)$ for arbitrary smooth periodic
$\chi:S^1\to\mathbb C^2$.
\end{proof}

\begin{remark}\label{rem:anchor-kernel}
The kernel of the anchor also selects the Killing number. For
$u\in\Gamma(\ker\rho_L)$ the operator $\nabla^S_u$ is
$C^\infty(M)$-linear, so along such directions the Killing equation
is algebraic. Here $u=q_a\mathcal X_a=\tfrac12dt$, and $q_aX_a=0$,
so that $\nabla^S_u=-\tfrac i4q_a\gamma_a$. The Killing equation
along $u$ therefore reads
\[
\bigl(\lambda+\tfrac i4\bigr)(q_a\gamma_a)\psi=0 .
\]
Since $(q_a\gamma_a)^2=|q|^2=1$, nonzero intrinsic Killing spinors
exist only for $\lambda=-i/4$. The sector $\lambda=+i/4$ is empty,
although the integrability condition \eqref{eq:spaceform-k-lambda}
with $k=1/4$ allows both signs.
\end{remark}

Let $\mathcal E^a$ denote the dual intrinsic coframe. Then
$d_L\mathcal E^1=-\mathcal E^2\wedge\mathcal E^3$ cyclically.
As in the sphere computation,
$\kappa=\mathcal E^3$ and
$\omega=\mathcal E^1\wedge\mathcal E^2$ are special KY and
CCKY forms with $c=1/4$, and for every $\chi(t)$,
\begin{equation}\label{eq:flux-product-actions}
\mathscr L^S_{\mathcal X_3}\chi=-\tfrac i2\gamma_3\chi,
\qquad
\mathscr L^{\mathrm{CCKY}}_\omega\chi=-\gamma_3\chi.
\end{equation}
These are nonzero for every nonzero solution. Theorem~\ref{prop:bridge}
applies to each $\mathcal X_a$ and identifies the first action with
the positive KD action.

The transverse freedom has a general elementary explanation:
if $d_Lf=0$, multiplication by $f$ preserves the intrinsic
Killing equation. Thus a nontransitive anchor need not give a
finite-dimensional Killing-spinor space. In this example,
$d_Lf(t)=0$, so $f(t)\mathcal X_a$ is also intrinsically Killing.
For nonconstant $f$, however,
\[
\mathcal L_{fX_a}g
=df\otimes X_a^{\flat_g}+X_a^{\flat_g}\otimes df
\]
is nonzero wherever $df\ne0$ and $X_a\ne0$.
Hence $f(t)\mathcal X_a$ is not an ambient generalized Killing
section in general: the additional hypothesis of
Theorem~\ref{prop:bridge} is substantive.

Nor are these intrinsic spinors ordinary geometric Killing spinors
of the product metric $g$. Its mixed curvature with the circle
direction is zero, whereas ordinary spinor integrability for
$\lambda\ne0$ would give
$R^S(X,Z)\psi=-2\lambda^2c(X)c(Z)\psi$ for orthonormal sphere
and circle directions $X,Z$. The Clifford product is invertible,
so no nonzero ordinary solution can satisfy that equation.
The intrinsic curvature is a curvature of $L$, not of $TM$.

\section{Isometry representations and symmetry brackets}\label{sec:algebra}

The intrinsic Killing sections form a real Lie algebra $\mathfrak k_0$
under $[\,,\,]_{L,H}$. Let $\mathfrak k_1$ be the underlying real
vector space of intrinsic Killing spinors with a fixed
$\lambda\in i\mathbb R$. The Hermitian pairing defines the real
symmetric current
\begin{equation}\label{eq:real-current}
h_L(V_{\psi,\phi},u)=\operatorname{Re}(\psi,c_L(u)\phi).
\end{equation}
By Proposition~\ref{prop:bilinears}, this current is a Killing
section. Consider the graded-antisymmetric operations
\begin{equation}\label{eq:degree-one-brackets}
[\mathcal K,\mathcal J]=[\mathcal K,\mathcal J]_{L,H},\qquad
[\mathcal K,\psi]=\mathscr L^S_{\mathcal K}\psi,\qquad
[\psi,\phi]=V_{\psi,\phi}.
\end{equation}

\begin{proposition}\label{prop:algebra}
The intrinsic Kosmann derivative defines a Lie algebra representation
of $\mathfrak k_0$ on $\mathfrak k_1$:
\begin{equation}\label{eq:isometry-representation}
[\mathscr L^S_{\mathcal K},\mathscr L^S_{\mathcal J}]
=\mathscr L^S_{[\mathcal K,\mathcal J]_{L,H}}.
\end{equation}
The current is also equivariant:
\begin{equation}\label{eq:bilinear-equivariance}
[\mathcal K,V_{\psi,\phi}]_{L,H}
=V_{\mathscr L^S_{\mathcal K}\psi,\phi}
 +V_{\psi,\mathscr L^S_{\mathcal K}\phi},
\end{equation}
Consequently the operations \eqref{eq:degree-one-brackets} satisfy
every graded Jacobi identity with at least one entry in
$\mathfrak k_0$.
\end{proposition}

\begin{proof}
Naturality of the spin lift of metric Lie algebroid automorphisms
gives \eqref{eq:isometry-representation}. One can see this locally by
lifting the commutator of the infinitesimal orthonormal-frame actions
to the spin principal bundle; the spin map is a Lie algebra
representation and has no central infinitesimal term. The flows
preserve the Levi--Civita connection by uniqueness of the Koszul
construction, which also justifies
\eqref{eq:Kosmann-connection-equivariance}. These are local
infinitesimal statements; completeness of the sections is not
required. Preservation of the Hermitian pairing and of the Clifford
action gives \eqref{eq:bilinear-equivariance}. The even--even--even
identity is the Jacobi identity of the Lie algebroid bracket, the
even--even--odd identity is \eqref{eq:isometry-representation}, and
the even--odd--odd identity is \eqref{eq:bilinear-equivariance}.
\end{proof}

The odd--odd--odd Jacobi identity additionally requires, in particular,
\begin{equation}\label{eq:odd-Jacobi-requirement}
\mathscr L^S_{V_{\psi,\psi}}\psi=0.
\end{equation}
It fails in the example of Section~\ref{sec:example}: for
$\psi=(1,0)^{\mathsf T}$ one has $V_{\psi,\psi}=E_3$ (transported
by $\iota_B$), while
\begin{equation}\label{eq:odd-Jacobi-failure}
\mathscr L^S_{V_{\psi,\psi}}\psi=-\frac{i}{2}\psi\ne0.
\end{equation}
The obstruction in the three-sphere example is a pure phase, and it
can be removed by a minimal enlargement of the even algebra.

\begin{proposition}\label{prop:phase-superalgebra}
For the graph Dirac structure on $SU(2)\simeq S^3$ in
Section~\ref{sec:example}, let $\mathfrak k_0$ denote its real Lie
algebra of intrinsic Killing sections and let $\mathfrak k_1$ be
the underlying real vector space of the Killing spinors with
$\lambda=-i/4$. Introduce an even generator $R$ and set
\[
\mathfrak g_{\bar 0}=\mathfrak k_0\oplus\mathbb R R,
\qquad \mathfrak g_{\bar 1}=\mathfrak k_1.
\]
Retain the even--even and even--odd brackets from
\eqref{eq:degree-one-brackets}, impose
\[
[\mathcal K,R]=0,\qquad [R,R]=0,\qquad [R,\psi]=i\psi,
\]
and define the symmetric odd bracket by
\begin{equation}\label{eq:phase-extended-odd-bracket}
[\psi,\phi]
 =V_{\psi,\phi}
 +\frac12\operatorname{Re}(\psi,\phi)R.
\end{equation}
Then $\mathfrak g=\mathfrak g_{\bar 0}\oplus\mathfrak g_{\bar 1}$,
with the graded-antisymmetric extension of these operations, is a
real Lie superalgebra.
\end{proposition}

\begin{proof}
The Killing spinors are constant two-component spinors in the
left-invariant spin frame. On this space, the intrinsic Kosmann
derivative along the orthonormal sections $E_a$ is
$\mathscr L^S_{E_a}=-\tfrac i2\gamma_a$.
For $\psi\in\mathfrak k_1$ let
$v_a=(\psi,\gamma_a\psi)\in\mathbb R$, so that
$V_{\psi,\psi}=v_aE_a$. The Pauli-matrix identity
$2\psi\psi^\dagger=\|\psi\|^2I+v_a\gamma_a$, where
$\|\psi\|^2=(\psi,\psi)$, implies $(v_a\gamma_a)\psi=\|\psi\|^2\psi$ and hence
\[
\mathscr L^S_{V_{\psi,\psi}}\psi
=-\frac i2\|\psi\|^2\psi.
\]
The modified odd bracket therefore obeys
\[
[[\psi,\psi],\psi]
=\mathscr L^S_{V_{\psi,\psi}}\psi
+\frac12\|\psi\|^2[R,\psi]=0.
\]
The odd--odd--odd Jacobi expression is a symmetric real-trilinear
map on $\mathfrak k_1$; its diagonal vanishing implies the full
identity by polarization over $\mathbb R$.

For $\mathcal K\in\mathfrak k_0$ the Kosmann derivative is compatible
with the Hermitian pairing,
$(\mathscr L^S_{\mathcal K}\psi,\phi)+(\psi,\mathscr L^S_{\mathcal K}\phi)
=\rho_L(\mathcal K)(\psi,\phi)$, and $(\psi,\phi)$ is constant on
$\mathfrak k_1$. Hence the $R$-component of
\eqref{eq:phase-extended-odd-bracket} is $\mathfrak k_0$-invariant,
and the Jacobi identities of Proposition~\ref{prop:algebra} involving
$\mathfrak k_0$ persist for the modified bracket. The phase action $[R,\psi]=i\psi$
commutes with every intrinsic Kosmann derivative. Moreover,
$V_{i\psi,\phi}+V_{\psi,i\phi}=0$ and
$\operatorname{Re}(i\psi,\phi)+
\operatorname{Re}(\psi,i\phi)=0$ by the Hermitian pairing.
Thus the Jacobi identity with $R$ and two odd entries also holds;
all remaining identities involving $R$ follow immediately from its
commutation with $\mathfrak k_0$ and its scalar action on
$\mathfrak k_1$. The displayed operations therefore close and
satisfy every graded Jacobi identity.
\end{proof}

The additional generator commutes with the even isometry algebra
but acts nontrivially on spinors; it is not central in the full
superalgebra. Proposition~\ref{prop:phase-superalgebra} provides
an explicit instance in which the Hermitian-current obstruction is
resolved by a phase extension. In rigid supersymmetric theories on
the three-sphere, the commutator of two supersymmetry transformations
likewise contains an $R$-rotation \cite{HamaHosomichiLee}. The
distinction between symmetry superalgebras and Lie superalgebras is
also emphasized in \cite[Sec.~VII]{AcikErtem}.

\begin{remark}\label{rem:flux-superalgebra}
The construction extends to the nontrivial-flux example of
Section~\ref{sec:nontrivial-flux}. There $\mathfrak k_1$ is the
underlying real space of $C^\infty(S^1,\mathbb C^2)$ and, as in
\eqref{eq:flux-product-actions},
$\mathscr L^S_{\mathcal X_a}\chi=-\tfrac i2\gamma_a\chi$ for
$a=1,2,3$. The current is $V_{\psi,\phi}=v_a\mathcal X_a$ with
$v_a=\operatorname{Re}(\psi,\gamma_a\phi)$ depending only on $t$;
since $d_Lv_a=0$, it is an intrinsic Killing section, and
$\mathscr L^S_{f\mathcal X_a}=f\,\mathscr L^S_{\mathcal X_a}$ whenever
$d_Lf=0$. Replace $\mathbb RR$ by the abelian algebra of generators
$R_f$, $f\in C^\infty(S^1)$, with $[\mathcal K,R_f]=0$ and
$[R_f,\psi]=if\psi$, and set
$[\psi,\phi]=V_{\psi,\phi}+\tfrac12R_{\operatorname{Re}(\psi,\phi)}$.
Because the anchor takes values in $TS^2$, $\rho_L(\mathcal K)f=0$ for
all $\mathcal K\in\mathfrak k_0$ and $f\in C^\infty(S^1)$; hence
$\mathscr L^S_{\mathcal K}$ commutes with every $R_f$ and preserves
$\operatorname{Re}(\psi,\phi)$. The proof of
Proposition~\ref{prop:phase-superalgebra} then applies pointwise in $t$
and yields an infinite-dimensional real Lie superalgebra on the
cohomologically nontrivial flux background.
\end{remark}

Whether comparable extensions exist for more general metric Dirac
structures remains open.

For higher-form extensions, the next problem is to identify a
natural bracket on a common space of KY/CKY forms compatible with
their spinorial actions. The integrability and bilinear results above
supply the geometric ingredients for this problem; closure and graded
Jacobi become additional, testable conditions.

\section{Conclusion}\label{sec:conclusion}
We have developed a spinorial symmetry framework for metric Dirac
structures in $H$-twisted Courant geometry. The generalized metric
provides an intrinsic Riemannian Lie-algebroid structure and an
isometry from $L$ to $V_+$. Our positive-spinor comparison theorem
shows that, for generalized Killing sections in $L$, the transported
Kosmann--Dorfman action agrees exactly with the intrinsic Kosmann
derivative. The cancellation of auxiliary connection differences
makes this comparison independent of the compatible generalized
connection used to express the lift.

The intrinsic Levi--Civita geometry supports a systematic treatment
of KY and CKY forms, their curvature integrability conditions, and
first-order symmetry operators on Killing spinors. Odd special KY
and even special closed CKY forms preserve the Killing equation for
$c=-4\lambda^2$; the Hermitian bilinears of Killing spinors with
imaginary $\lambda$ furnish these forms directly. Together with the
positive-spinor intertwining, these results connect generalized
isometries to higher-form symmetries without conflating the ambient
and intrinsic Clifford modules.

The explicit examples display two distinct aspects of this theory.
The graph construction on $S^3$ verifies the correspondence in an
exact-flux setting. More significantly, the $S^2\times S^1$ model
has a nontrivial $H$-flux class and noninvertible anchor, while
admitting globally defined spinors and nonvanishing KY/CKY symmetry
actions. Its Killing-spinor space
$C^\infty(S^1,\mathbb C^2)$ illustrates a characteristic feature
of nontransitive Lie-algebroid geometry: transverse parameters can
survive the intrinsic Killing equation even when the corresponding
ordinary metric admits no Killing spinors with that Killing number,
while the kernel of the anchor selects a single Killing number.
Intrinsic Killing sections act on Killing spinors by a Lie algebra
representation, and the Hermitian currents transform equivariantly.
In the $S^3$ example, the current bracket has a pure-phase
odd-Jacobi obstruction, which we remove explicitly by adjoining one
even phase generator; the resulting brackets satisfy all graded
Jacobi identities and furnish a real Lie superalgebra. A family of
such generators parametrized by $C^\infty(S^1)$ does the same on the
$S^2\times S^1$ flux background.

Several concrete research directions follow. One is to understand
Killing-spinor families and symmetry algebras for Dirac structures
with singular or nontransitive anchors, including their dependence
on leafwise geometry and transverse data. Another is to study the
transport of positive-spinor symmetries under Courant-algebroid
isomorphisms and topological T-duality
\cite{BouwknegtEvslinMathai,GarciaFernandezRicci}, beyond the worked
flux background here. A third is to extend the KY/CKY construction to
specified torsionful Lie-algebroid connections and to investigate
higher-form brackets compatible with the induced spinorial actions.
The phase-extension mechanism in Proposition~\ref{prop:phase-superalgebra}
also suggests examining geometric conditions under which related
even generators restore graded Jacobi identities beyond the
two examples treated here.
The present results provide a controlled starting point for these
extensions and show that Dirac structures offer a productive setting
for hidden symmetries in flux geometry.

\begin{acknowledgments}

This study was supported by Scientific and Technological Research Council of T\"urkiye (T\"UB\.ITAK) under the Grant Number 123F261. The authors thank T\"UB\.ITAK for its support.

\end{acknowledgments}

\end{document}